\documentclass[aps,twocolumn,superscriptaddress,showpacs,floats]{revtex4-1}
\usepackage{amsmath}
\usepackage{amsfonts}
\usepackage{amssymb}
\usepackage{euscript}
\usepackage{bm}
\usepackage[dvips]{graphicx}

\begin{document}

\title{The 17/5 spectrum of the Kelvin-wave cascade}

\author{Evgeny Kozik}
 \affiliation{Institute for Theoretical Physics, ETH Zurich, CH-8093 Zurich, Switzerland}
\author{Boris Svistunov}
\affiliation{Department of Physics, University of Massachusetts,
Amherst, MA 01003} \affiliation{Russian Research Center
``Kurchatov Institute'', 123182 Moscow, Russia}

\begin{abstract}
Direct numeric simulation of the Biot-Savart equation readily resolves the 17/5 spectrum of the Kelvin-wave cascade
from the 11/3 spectrum of the non-local (in the wavenumber space) cascade scenario by L'vov and Nazarenko. This result is a clear-cut visualisation of the unphysical nature of the 11/3 solution, which was
established earlier on the grounds of symmetry. 
\end{abstract}


%
%
%
%
%

\maketitle

 A cascade of Kelvin waves (kelvons)---distortion waves on quantized vortex filaments---plays a crucial role in the  decay of superfluid turbulence at $T=0$ by providing the  mechanism of energy transfer \cite{Sv95,Vinen_2003, KS_04,KS_05} to shorter length scales, where the dissipation due to phonon emission becomes efficient \cite{Vinen2000, KS_05_vortex_phonon} (for a review of the theory of superfluid turbulence 
 in the  $T\to 0$ limit, see Ref.~\cite{KS_JLTP}).  Our previous analytic \cite{KS_04} and numeric \cite{KS_05}  studies of the Kelvin-wave cascade in the regime of weak turbulence
 revealed the spectrum 
\begin{equation}
n_k \, \propto\, k^{-17/5} 
\label{17_5}
\end{equation}
for the kelvon occupation numbers. Since then this result---implying  the locality of the three-kelvon elastic collisions in the wave number space---was considered well-established. However, recently,   Laurie,  L'vov, Nazarenko, and
Rudenko claimed the three-kelvon (a.k.a. six-wave) collisions to be essentially \textit{non-local} \cite{LN_nonlocal}.  An alternative  theory of the Kelvin-wave cascade was then put forward by L'vov and Nazarenko \cite{LN},
with the spectrum 
\begin{equation}
n_k \, \propto\, k^{-11/3}  \qquad \mbox{(L'vov and Nazarenko)}.
\label{LN}
\end{equation}
Our subsequent analysis of the problem \cite{KS_geom_sym} brought us to the conclusion that the locality of the Kelvin-wave cascade can be rigorously proven on the basis of geometric symmetries alone. Moreover, 
we argued that the spectrum (\ref{LN}) is characteristic of (uncontrollable) approximations, as well as errors, violating the so-called  tilt symmetry. We have also remarked that proposed in Ref.~\cite{LN_nonlocal} 
`additional' terms of the collisional integral---absent in our analysis \cite{KS_JLTP}---are  likely to be the source of the mistake. The reaction of the community to the argument of Ref.~\cite{KS_geom_sym} remains essentially reserved 
\cite{reaction}. Therefore, we find it instructive to visualize the argument by first-principle direct simulations. 

Vortex dynamics at $T=0$ is described by the (asymptotically exact in the hydrodynamic limit) Biot-Savart equation (BSE). The BSE is notoriously difficult to integrate numerically due to the essentially non-local (in real space) interaction between vortex-line elements. Moreover, it is this interaction that accounts for kelvon kinetics in the weak-turbulence regime in question. To efficiently simulate the BSE,  in our Ref.~\cite{KS_05}, we developed the so-called scale-separation scheme, which reduces the simulation cost of BSE to that of a local model. Back in 2005, the efficiency of the scheme was sufficient to unambiguously resolve the difference 
between the spectrum $k^{-17/5}$ and the previously conjectured $k^{-3}$, using a standard desktop computer. Given that the difference between $11/3$ and $17/5$ is of the same order as the difference between $17/5$ and $3$, it is expected that the numeric protocol of Ref.~\cite{KS_05} should allow one to readily resolve $17/5$ from $11/3$ with a modern laptop.

\begin{figure}[htb]
\includegraphics[width = 0.97\columnwidth,keepaspectratio=true]{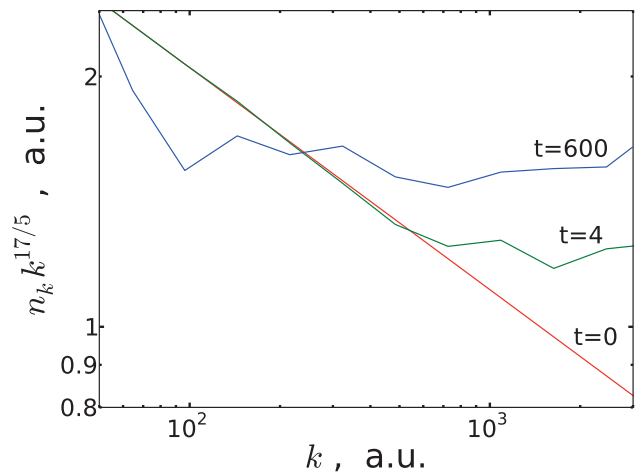}
\caption{(Color online) The time evolution of the Kelvin-wave spectrum plotted as $n_k k^{17/5}$. The initial condition ($t=0$) corresponds to the spectrum (\ref{LN}). As the time $t$ (measured in the units of the period of the fastest mode in the system) progresses, the spectrum (\ref{LN}) is transformed by a wave propagating from high wavenumbers to the smaller ones into the expected $n_k \propto k^{-17/5}$ spectrum.}
\label{f1}
\end{figure}

The results of our simulation are presented in Fig.~\ref{f1}. For convenience of visual inspection we plot $n_k k^{17/5}$ so that the spectrum (\ref{17_5}) corresponds to a horizontal line. We use the same routine as in Ref.~\cite{KS_05}, with the only difference that the initial condition (labelled with $t=0$) now corresponds to the spectrum (\ref{LN}) rather than $n_k\propto k^{-3}$. We show the evolution of the spectrum at two instants, $t=4$ and $t=600$ (in the units of the period of the fastest mode in the system). The spectra follow from time averaging of the wavenumber distribution over intervals $\Delta t \lesssim t$ suggested by the quasi-steady-state nature of a cascade. In order to filter the inherent noise between neighbouring wavenumbers, we obtain $n_k$ after averaging over the wavenumber range $[1.5^{-1/2}k,\,   1.5^{1/2} k ]$. The simulation took 12 hours on a laptop with a 2.4 GHz processor. We clearly see that the distribution $n_k \, \propto\, k^{-11/3}$ is being re-structured into  $n_k \, \propto\, k^{-17/5}$, the transient having the form of a wave propagating from large to smaller wavenumbers, in a direct analogy with the picture seen in Ref.~\cite{KS_05}. It is important to note that during the transient the amplitude
of the $k^{-17/5}$ tail (the height of the horizontal piece of the curve) {\it increases}. This is a qualitative feature immediately ruling out the spectrum~(\ref{LN}).

In conclusion, we make two remarks of a more general character.
First, we see that  there is no need for developing approximate local (in real space) models  since, in numerical simulations of kelvons on individual vortex lines, the full spatially non-local model  can be efficiently solved at the expense of a local one (and when the coupling between separate lines is important, no local model is applicable in principle). Second, the efficiency of the scale-separation scheme calls for the implementation of  tree codes
in simulations of  vortex tangles.

The authors acknowledge the hospitality of the Les Houches School of Physics (E.K.) and the University of Ghent (B.S.), where this work was completed.


\begin{thebibliography}{99}


 \bibitem{Sv95} B.V. Svistunov, Phys. Rev. B {\bf 52}, 3647 (1995).
 
 \bibitem{Vinen_2003} W.F. Vinen, M. Tsubota and A. Mitani, Phys.
Rev. Lett. {\bf 91}, 135301 (2003).

\bibitem{KS_04} E. Kozik and B. Svistunov, Phys. Rev. Lett. {\bf 92}, 035301 (2004).

\bibitem{KS_05} E. Kozik and B. Svistunov, Phys. Rev. Lett.  {\bf 94}, 025301 (2005).
 
 \bibitem{Vinen2000} W.F. Vinen, Phys. Rev. B  {\bf 61}, 1410 (2000).

\bibitem{KS_05_vortex_phonon} E. Kozik and B. Svistunov, Phys. Rev. B  {\bf 72}, 172505 (2005). 
 
 \bibitem{KS_JLTP} E. Kozik and B. Svistunov, J. Low Temp. Phys.
\textbf{156}, 215 (2009).



\bibitem{LN_nonlocal} J. Laurie, V. S. L'vov, S. Nazarenko, and O. Rudenko, 
Phys. Rev. B \textbf{81}, 104526 (2010).


\bibitem{LN} V.S. L'vov and S. Nazarenko, Pis'ma v ZhETF \textbf{91}, 464
(2010). The reader should not be mislead by the word
`curvature' mistakenly used there instead of `angle'.

 \bibitem{KS_geom_sym} E. Kozik and B. Svistunov, arXiv:1006.0506.
 
 \bibitem {reaction} We refer to conversations with V. Lebedev and G. Volovik, to reports of the referees on Ref.~\cite{KS_geom_sym}, as well as the
 attempt by V. Lebedev and V. L'vov (see arXiv:1005.4575) to construct counter-arguments to our proof (we note in passing that the inconsistency of these counter-arguments is
 explained  in our arXiv:1006.1789; Lebedev, L'vov, and Nazarenko do not agree with this explanation [arXiv:1007.3191]).

\end{thebibliography}
\end{document}